\documentclass[amsmath,amssymb,nofootinbib,notitlepage,prd,superscriptaddress,twocolumn]{revtex4-2}
\usepackage[T1]{fontenc}
\usepackage[usenames,dvipsnames]{xcolor}
\usepackage{graphicx,hyperref,orcidlink,float}
\hypersetup{colorlinks=true,citecolor=RoyalBlue,linkcolor=RoyalBlue,urlcolor=RoyalBlue}

\usepackage{appendix}
\usepackage{graphicx}

\usepackage{tikz}
\usepackage{pgfplots}
\usepackage{dcolumn}
\usepackage{bm}
\usepackage{amsmath}

\usepackage[mathlines]{lineno}
\usepackage{xcolor}
\usepackage{slashed}

\newcommand{\KM}[1]{\textcolor{purple}{KM: #1}}

\usepackage[commentmarkup=uwave,commandnameprefix=ifneeded]{changes}

\usepackage[normalem]{ulem}

\usepackage[normalem]{ulem}

\pgfplotsset{compat=1.18}
\begin{document}

\title{Nonlinear Effects In Black Hole Ringdown From Scattering Experiments I: spin and initial data dependence of quadratic mode coupling}

\newcommand{\Caltech}{\affiliation{Theoretical Astrophysics, Walter Burke
  Institute for Theoretical Physics, California Institute of Technology,
  Pasadena, California 91125, USA}}
\newcommand{\Perimeter}{\affiliation{Perimeter Institute for Theoretical Physics, 31 Caroline Streeth North, Waterloo, Onatrio NSL 2Y5, Canada}}
\newcommand{\PGI}{\affiliation{Princeton Gravity Initiative, Princeton University, Princeton, New Jersey, 08544, USA}}
\newcommand{\Cornell}{\affiliation{Cornell Center for Astrophysics and Planetary Science, Cornell University, Ithaca, New York 14853, USA}}
\newcommand{\AEI}{\affiliation{Max Planck Institute for Gravitational Physics (Albert Einstein Institute), Am M\"uhlenberg 1, D-14476 Potsdam, Germany}}
\newcommand{\Princeton}{\affiliation{Department of Physics, Princeton University, Jadwin Hall, Washington Road, New Jersey, 08544, USA}}
\newcommand{\UIUC}{\affiliation{Illinois Center for Advanced Studies of the Universe \& Department of Physics, University of Illinois at Urbana-Champaign, Urbana, IL 61801, USA}}
\newcommand{\Oberlin}{\affiliation{Department of Physics and Astronomy, Oberlin College}}

\author{Hengrui Zhu\orcidlink{0000-0001-9027-4184}}
\email{hengrui.zhu@princeton.edu}
\Princeton
\PGI

\author{Justin L. Ripley\orcidlink{0000-0001-7192-0021}}
\UIUC

\author{Frans Pretorius}
\Princeton
\PGI

\author{Sizheng Ma\orcidlink{0000-0002-4645-453X}}
\Perimeter

\author{Keefe Mitman\orcidlink{0000-0003-0276-3856}}
\Caltech

\author{Robert~Owen}
\Oberlin

\newcommand{\CornellPhysics}{\affiliation{Department of Physics,
Cornell University, Ithaca,
    NY, 14853, USA}}

  \newcommand{\MaxPlanck}{\affiliation{Max Planck Institute for
Gravitational
      Physics (Albert Einstein Institute), Am M{\"u}hlenberg 1, D-14476
Potsdam,
      Germany}}

  \author{Michael Boyle \orcidlink{0000-0002-5075-5116}} \Cornell
  \author{Yitian Chen \orcidlink{0000-0002-8664-9702}} \Cornell
  \author{Nils Deppe \orcidlink{0000-0003-4557-4115}} \CornellPhysics
\Cornell
  \author{Lawrence E.~Kidder \orcidlink{0000-0001-5392-7342}} \Cornell
  \author{Jordan Moxon \orcidlink{0000-0001-9891-8677}} \Caltech
  \author{Kyle C. Nelli \orcidlink{0000-0003-2426-8768}} \Caltech
  \author{Harald P.~Pfeiffer \orcidlink{0000-0001-9288-519X}} \MaxPlanck
  \author{Mark A.~Scheel \orcidlink{0000-0001-6656-9134}} \Caltech
  \author{William Throwe \orcidlink{0000-0001-5059-4378}} \Cornell
  \author{Nils L.~Vu \orcidlink{0000-0002-5767-3949}} \Caltech

\date{\today}

\begin{abstract}
We investigate quadratic quasinormal mode coupling in black hole spacetime through numerical simulations of single perturbed black holes using both numerical relativity and second-order black hole perturbation theory. 
Focusing on the dominant $\ell=|m|=2$ quadrupolar modes, we find good agreement (within $\sim10\%$) between these approaches, with discrepancies attributed to truncation error and uncertainties from mode fitting.
Our results align with earlier studies extracting the coupling coefficients from select binary black hole merger simulations, showing consistency for the same remnant spins. 
Notably, the coupling coefficient is insensitive to a diverse range of initial data, including configurations that led to a significant (up to $5\%$) increase in the remnant black hole mass.
These findings present opportunities for testing the nonlinear dynamics of general relativity with ground-based gravitational wave observatories. 
Lastly, we provide evidence of a bifurcation in coupling coefficients between counter-rotating and co-rotating quasinormal modes as black hole spin increases.
\end{abstract}

\maketitle

\section{Introduction}
According to the theory of general relativity, the merger of two black holes results in a highly distorted remnant that quickly rings down to a stationary Kerr black hole~\cite{Teukolsky:1973ha, Owen:2009sb,Buonanno:2006ui}. 
The gravitational wave signal during ringdown has been shown to be well-described by black hole perturbation theory~\cite{Vishveshwara:1970zz,Press:1971wr,Price:1994pm,Buonanno:2006ui, Berti:2007fi, LIGOScientific:2016vlm,Giesler:2019uxc,Li:2021wgz}. 
To linear order, black hole perturbation theory is governed by the spin-$\pm2$ Teukolsky equation, which is a separable partial differential equation governing a single complex scalar field representing the two gravitational wave polarizations~\cite{Teukolsky:1973ha,Teukolsky:1974yv}. 
The ringdown is expected to be dominated by quasinormal modes (QNMs):  damped sinusoids that are eigen-solutions of the Teukolsky operator~\cite{Leaver:1985ax,Berti:2009kk,Konoplya:2011qq}. 
The complex frequencies of these QNMs are determined by the mass and spin of the remnant black hole, thereby offering a unique window to probe its properties and test the (linearized) dynamics of general relativity~\cite{Detweiler:1980gk,Dreyer:2003bv,Berti:2005ys,Isi:2020tac, Carullo:2019flw, Isi:2021iql}. 

It has also been shown that to second order in perturbation theory, the dynamics of a Kerr black hole is governed by the same Teukolsky operator $\mathcal{T}$, but now with a source term $\mathcal{S}^{(1)}$ quadratic in the linear order curvature perturbation $\psi_4^{(1)}$~\cite{Campanelli:1998jv}:
\begin{equation}\label{eq:sec}
    \mathcal{T}\psi_4^{(2)} = \mathcal{S}^{(1)}\sim \left(\psi_4^{(1)}\right)^2~,
\end{equation}
where $\psi_4^{(2)}$ gives the second order correction to the curvature perturbation.
When the linear perturbation contains QNMs, the equation above predicts QQNMs~\cite{Gleiser:1995gx,Brizuela:2009qd,Ripley:2020xby,Loutrel:2020wbw,Ioka:2007ak,Nakano:2007cj,Pazos:2010xf,London:2014cma,Lagos:2022otp}. These QQNMs represent particular solutions of the sourced Teukolsky equation, with complex frequencies as the sum of those from the driving modes. 
Additionally, their amplitudes exhibit a quadratic dependence on those of the driving modes.

We now introduce our QNM conventions for clarity. As in Ref~\cite{Isi:2021iql}, we label the complex frequency and amplitude of a QNM by $\omega_{plmn}$ and $A_{plmn}$, where $l$ and $m$ are the quantum numbers associated with the QNM angular eigenfunctions (spin-weighted spheroidal harmonics), $n$ is the overtone index associated with the QNM radial eigenfunctions, and $p\in\{\pm\}$ indexes the prograde $(+)$ and retrograde $(-)$ modes. For prograde modes, the phase front rotates in the same angular direction as the horizon, while the phase front of of retrograde modes counter-rotates. 

We further write the frequency and amplitude for a QQNM, driven quadratically by two linear QNMs (primed and unprimed), as follows:
\begin{align}
    \omega_{p l m n\times p^\prime l^\prime m^\prime n^\prime}&=\omega_{p l m n}+\omega_{p^\prime l^\prime m^\prime n^\prime}~~~~\mathrm{and}\\
    A_{p l m n\times p^\prime l^\prime m^\prime n^\prime} &= \mathcal{R}_{p l m n\times p^\prime l^\prime m^\prime n^\prime}A_{p l m n}A_{p^\prime l^\prime m^\prime n^\prime}~,\label{eq:quadratic}
\end{align}
where $\mathcal{R}_{p l m n \times p^\prime l^\prime m^\prime n^\prime}$ is a proportionality constant that is determined by the second order dynamics of General Relativity. 
We refer to $\mathcal{R}_{p l m n\times p^\prime l^\prime m^\prime n^\prime}$ as the quadratic coupling coefficient, which is the main focus of this paper. 

For black hole remnants formed from binary coalescence, one expects the quadrupolar ${l=|m|=2}$ modes to be maximally excited in non-precessing, quasi-circular black hole binaries with close-to-unity mass ratio~\cite{Buonanno:2006ui,Berti:2007fi,Giesler:2019uxc,Kamaretsos:2012bs}. 
Therefore, the quadratic coupling of the prograde ${l=|m|=2}$ fundamental mode is of great interest, 
as an astrophysical measurement of the QQNM would directly probe the nonlinear dynamics of Kerr black holes as predicted by general relativity. 

Recent works have identified QQNMs from the ringdown signal in numerical relativity simulations of binary black hole mergers \cite{London:2014cma,Ma:2022wpv,Mitman:2022qdl,Cheung:2022rbm,cheung2023extracting,Khera:2023lnc}. 
In particular, Refs~\cite{Mitman:2022qdl,Cheung:2022rbm} found that the quadratic coupling coefficient for the prograde $l=m=2$ fundamental mode was between 0.14 and 0.19 for the black hole remnants with dimensionless spin ranging from $\chi=0.5-0.7$, formed from non-precessing, quasi-circular binaries with mass ratio ranging from 1 to 8.
Semi-analytical results using the WKB approximation for a Schwarzschild black hole, and calculations using the Kerr/CFT correspondence also produce numbers in this range \cite{Perrone:2023jzq,Kehagias:2023ctr}. 
Finally, an ongoing semi-analytical work without these approximations also finds a coupling coefficient consistent with this for a Schwarzschild black hole~\cite{Ma_in_prep}.

In this paper, we present results from a series of numerical experiments for a single perturbed black hole, both by solving the fully nonlinear Einstein equations (through numerical relativity) and the second order Teukolsky equation, in an effort to gain a better understanding of the spin and initial data dependence of quadratic mode coupling. 
In brief, we find that: 
\begin{enumerate}
    \item The quadratic coefficient measured from full numerical relativity simulations is insensitive to a wide class of initial conditions within fitting/truncation error, even when the backreaction to the background is substantial. 
    \item The quadratic coefficients measured using second order perturbation theory are within 10\% of those obtained from numerical relativity simulations, for all black hole spins we considered. This discrepancy is within the estimated uncertainties for high spin, but outside of it at low spin.\footnote{We cannot rule out an as-of-yet unidentified error, or unaccounted for systematic in the rather vast pipeline going from multiple aspects of theory to mode extraction.} 
    \item The quadratic coefficient monotonically decreases with spin up to $\chi=0.99$, but not to zero. 
    \item The quadratic coupling coefficient for the retrograde mode bifurcates from that for the prograde as the spin of the black hole increases.
\end{enumerate}

Using the same second order perturbation theory code as we do here~\cite{Loutrel:2020wbw,Ripley:2020xby}, and similar scattering initial data, Ref.~\cite{Redondo-Yuste:2023seq} also explored the prograde quadratic coupling coefficient as a function of spin, and some aspects of the sensitivity to initial data. 
Their results however disagree with ours, and with those measured from the merger simulations mentioned above, by more than a factor of 2. 
They further find that the quadratic coupling coefficient decreases to zero as a function of the dimensionless spin of the background, again in disagreement with our results. 

This paper is organized as follows: in Sec.~\ref{sec:num} we describe our numerical setup for simulating a perturbed black hole with both numerical relativity and second order perturbation theory, and in Sec.~\ref{sec:ana} we present our procedure for extracting the quadratic coefficient.
In Sec.~\ref{sec:res} we show our results regarding the spin and initial data dependence of the quadratic mode coupling, and the coupling between retrograde QNMs. Lastly, we conclude in Sec.~\ref{sec:conclude}. 
In the supplemental material, we discuss the impact of tetrad convention on the quadratic coefficient in Sec.~I, and numerical convergence in Sec.~II.
In Sec III we assess the impact of wave extraction method. Lastly we present additional evidence regarding the coupling between retrograde modes in Sec.~IV.

\section{Numerical Setup}\label{sec:num}

\subsection{Evolution in full numerical relativity}
To extract the quadratic coupling coefficient from the fully nonlinear Einstein equations, we numerically simulate a single perturbed 
Kerr black hole with the Spectral Einstein Code (\texttt{SpEC}), using the generalized harmonic formulation \cite{Lindblom:2005qh,Scheel:2006gg,Pretorius:2004jg}. 
To assess the initial data dependence, we use three different configurations for the initial perturbation, representing an ingoing pulse from large distance, an outgoing pulse from the vicinity of the black hole horizon, or a pulse co-rotating with the black hole horizon. 
We perform the simulations in Spherical Kerr-Schild (SKS) coordinates \cite{Chen:2021rtb}, 
where the coordinate shape of the horizons are round spheres regardless of the black hole spin, 
simplifying the excision procedure. 

The angular sector of the metric perturbation is chosen to be a pure-spin tensor spherical harmonic. 
Following the convention in Ref.~\cite{Martel:2005ir} for the angular components, we have
\begin{align}
    h_{ij}(r,\theta,\phi) &= R(r) Y^{lm}_{ij}(\theta,\phi)~,
\end{align}
where $R(r)$ is the radial profile of the gravitational perturbation, and we set the radial tensorial components to zero, that is $Y_{ri}^{lm}=0$ for $i\in\{r,\theta,\phi\}$. 
The time derivatives differ for the three class of initial conditions:
\begin{enumerate}
    \item ingoing $(+)$ and outgoing $(-)$ pulse initial data: 
    \begin{equation}\label{eq:io}
        \partial_t h_{ij}(r,\theta,\phi) = \pm\frac{d R(r)}{dr} Y^{lm}_{ij}~,
    \end{equation}
    \item rotating pulse initial data: 
    \begin{equation}\label{eq:rot}
        \partial_t h_{ij}(r,\theta,\phi) = -\Omega R(r) \partial_\phi Y^{lm}_{ij},
    \end{equation}
\end{enumerate}
where $\Omega$ is the angular frequency of the pulse, and we restrict $R(r)$ to be a Gaussian of the form:
\begin{equation}\label{eq:id_radial}
    R(r,w,r_0) = \exp\left(-\frac{(r-r_0)^2}{w^2}\right)~.
\end{equation}

We choose $r_0=30 M$ for the ingoing pulse (scattering) initial condition, and $r_0=3 M$ otherwise. 
For the rotating pulse, we set $\Omega=\Re\{\omega_{+220}\}$ so that the pulse co-rotates with the horizon; we refer to this initial condition as the prograde pulse for the remainder of the paper.

For the angular sector, in this work we restrict our attention to initial data that only contains a perturbation with ${l=m=2}$ at linear order. Quadratic coupling between other angular modes can also be investigated based on the formalism outlined here.

The approximate spatial metric and its derivative on the Cauchy slice, to first order in the perturbation amplitude A is then given by
\begin{subequations}
\label{eq:spatial-initial-data-NR}
\begin{align}
    g_{ij}(r,\theta,\phi) 
    &\approx g^{SKS}_{ij}(r,\theta,\phi) + A h_{ij}(r,\theta,\phi)~, \label{eq:id1}
    \\
    \partial_t g_{ij}(r,\theta,\phi) &\approx A \partial_t h_{ij}(r,\theta,\phi)~.  \label{eq:id2}  
\end{align}
\end{subequations}
To ensure that our initial data satisfies the Hamiltonian and momentum constraint equations, we set Eq.~\eqref{eq:spatial-initial-data-NR} to be the conformal metric and its time derivative, and solve for the shift and the conformal factor using the conformal-thin-sandwich 
solver implemented in \textsc{SpEC}, according to the procedure outlined in Ref.~\cite{Pfeiffer:2004qz}.

For a given simulation, we calculate the waveform at future null infinity using the Cauchy-Characteristic Evolution (CCE) module implemented in \textsc{SpECTRE}~\cite{Bishop:1996gt,Reisswig:2009rx,Bishop:2016lgv,Barkett:2019uae,Moxon:2020gha,Moxon:2021gbv,spectrecode}. 
Then, we perform a BMS transformation to transform the resulting waveform to the superrest frame of the remnant black hole, using the procedure outlined in Refs.~\cite{MaganaZertuche:2021syq,Mitman:2022kwt} and the code \textsc{scri}~\cite{mike_boyle_2020_4041972,Boyle:2015nqa}.

\subsection{Evolution with second order perturbation theory}\label{sec_evo_sec}

To extract the quadratic coupling coefficient from the second order black hole perturbation theory, we use the code developed in Ref.~\cite{Ripley:2020xby,justin_ripley_2023_8215577} to solve the second order Teukolsky equation (c.f. Eq.~\eqref{eq:sec}).
The code adapts a hyperboloidal coordinate, which connect the black hole horizon to future null infinity in a single smooth slice~\cite{zenginouglu2008hyperboloidal,PanossoMacedo:2019npm}. We use the waveform at the future null infinity to be fully consistency with numerical relativity.

The choice of initial conditions for the second order code is limited, as constraint-satisfying initial data on the same Cauchy slice $T=T_0$ for the first-order perturbation is not yet available. 
Instead, as in Ref.~\cite{Ripley:2020xby}, we set up a radially ingoing pulse with compact support. 
Similar to the nonlinear setup, we specify the angular sector of the perturbation to be the ${\ell=m=2}$ spin-weighted spherical harmonics for the first-order perturbation.
To the causal future of the trailing edge for this pulse, reconstructing the first order metric needed for the second order source function does not require solving a Cauchy constraint problem. 
Consequently, integration of the second order Teukolsky equation starts at $T=T_1 > T_0$, where $T_1$ is entirely within the region where a constraint-satisfying source function is known. 
This means from the perspective of the full perturbation up to second order, the {\em effective} initial data is at $T=T_1$, with that for the first order perturbation differing from the functional forms we specify at $T=T_0$ by forward propagation with the Teukolsky operator to $T=T_1$.

With that said, regardless of the spin of the background black hole, at $T=T_0$ we choose pulse widths of $1.8$, $2.8$, and $3.8$ M, with the inner edge of each pulse fixed at $2.2$ M.
We refer readers to Ref.~\cite{Ripley:2020xby} for further details regarding the initial conditions. 

\begin{figure*}[h!tbp]%
    \includegraphics[width=2\columnwidth]{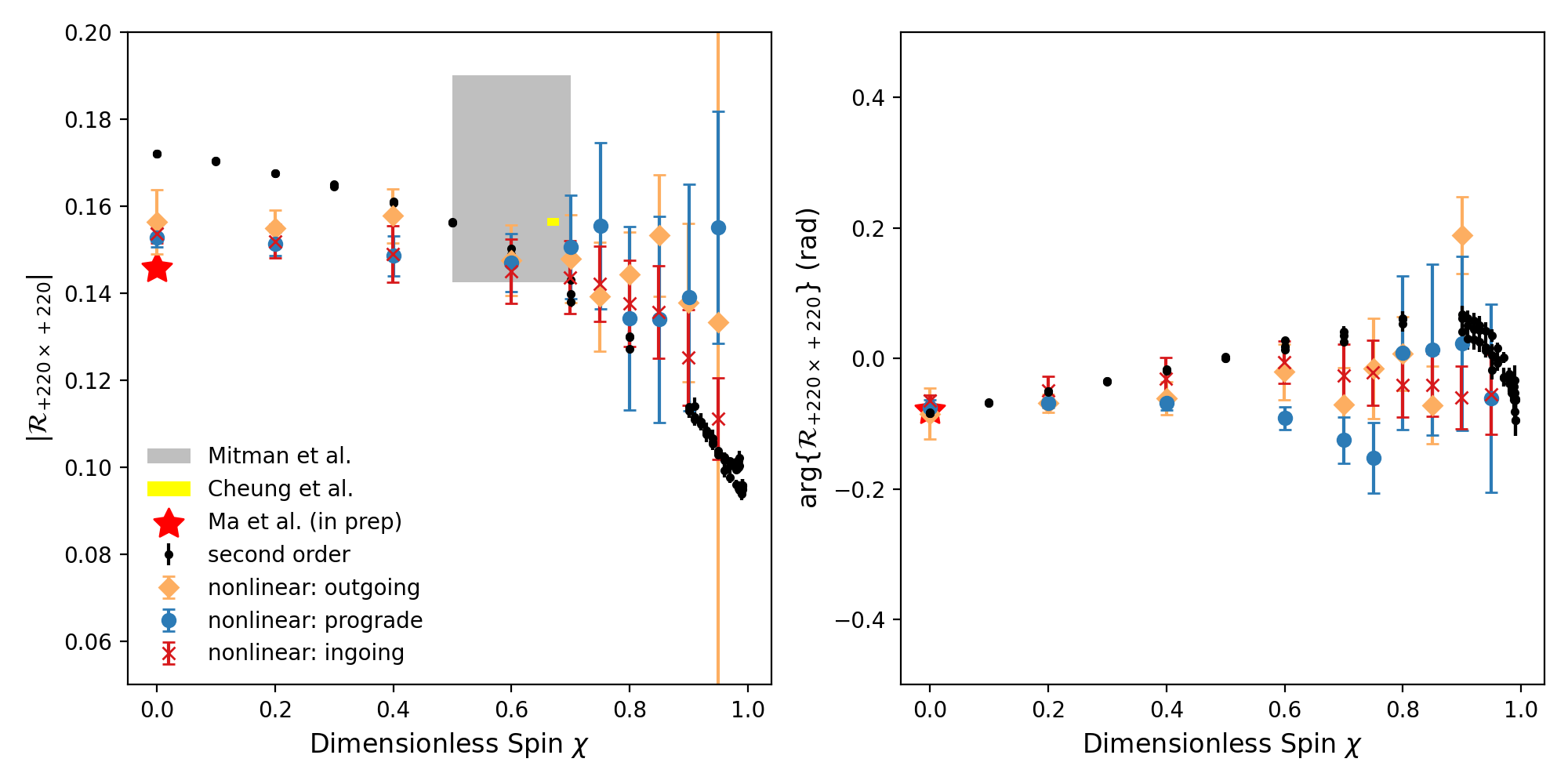}
    \vspace{0.3cm}
    \centering
    \caption{
    Amplitude (left) and phase (right) of the quadratic coupling coefficient for the prograde ${l=m=2}$ fundamental mode as a function of black hole spin from the second order perturbation theory (black) and full numerical relativity (colored) codes. 
    Results from numerical relativity are color-coded by the class of initial condition (Eqs.~\ref{eq:io}-\ref{eq:rot}); the width of each pulse is fixed to be $w=2M$ (Eq.~\ref{eq:id_radial}), with an amplitude such that the change in black hole mass and spin is less than $0.1\%$.
    The multiple black dots at the same spin represent initial conditions with the three different pulse widths used in  the second order code (see Sec.~\ref{sec_evo_sec}).
    Previous numerical results are shown as boxes with the reported range of spin and uncertainties, where ``Mitman et. al.'' refers to \cite{Mitman:2022qdl}, and ``Cheung et. al.'' refers to \cite{Cheung:2022rbm}; and ``Ma et al.'', an ongoing semi-analytical calculation~\cite{Ma_in_prep}, is shown as the red star.
    The deviation at low spin is unexpected---see the text for a discussion.
    }
    \label{fig:1}
\end{figure*}
\section{Analysis}\label{sec:ana}
We perform a linear least-squares fit to the resulting waveforms in terms of the Weyl scalar $\psi_4$ at future null infinity with linear sums of damped sinusoids, for both the nonlinear and second order code. 
For the nonlinear simulations, the mass of the black hole may change, so we rescale the waveform time and normalize the waveform by the remnant mass:
\begin{subequations}
\begin{align}
    t 
    &\to t/M_{\rm{rem}} 
    \ ,
    \\
    \psi_4 
    &\to 
    r\psi_4 M_{\rm{rem}}
    \ .
\end{align}
\end{subequations}
In previous studies regarding binary black holes, the rescaling is done using the initial Christodoulou mass of the binary $M_{\rm{init}}$, as in Refs.~\cite{Mitman:2022qdl,Cheung:2022rbm}, which results in a bias in the measured quadratic coefficient (in strain) by a factor of $M_{\rm{rem}}/M_{\rm{init}}$. This factor is typically 0.95 for near equal-mass binaries. Therefore, we multiplied numbers quoted in Ref.~\cite{Mitman:2022qdl,Cheung:2022rbm} by 0.95.

Furthermore, as we discuss in the supplemental material, different tetrad conventions can lead to different numerical values of the quadratic coefficient. 
Here, we normalize the waveform quantities based on the tetrad and coordinates conventions adopted in the \textsc{SpEC} code \cite{Iozzo:2020jcu}.
Lastly, while we fit $\psi_4$ to obtain the amplitudes for the QNMs, we convert the $\psi_4$ amplitudes to that of the gravitational strain for comparing with previous results. We use that $\psi_4=-\ddot{h}$ at the future null infinity, so for QNMs with the form in Eq.~\ref{eq:mode}, the amplitude in strain is given by $A^h_{plmn} = A^{\psi_4}_{plmn}/\omega_{plmn}^2$. 

For the $m=2$ modes (or the linear sector), we use the ansatz:
\begin{align}\label{eq:mode}
    &\psi_{4;lm}(t) = \nonumber \\ &\sum_{l,m,n,p\in\{\pm\}}A_{plmn} \sum_{l'}c_{pll'mn}~\exp\{-i\omega_{pl'mn} t\}~,
\end{align}
where $c_{pll'mn}$ is the spherical-spheroidal mixing coefficient, $\omega_{pl'mn}$ is the frequency of the linear quasinormal modes (calculated using the \texttt{qnm} package \cite{Stein:2019mop}), and $A_{plmn}$ is the amplitude for the QNM. 

For $m=2$, we simultaneously fit for the $l=2$ and $l=3$ modes, each with the prograde and retrograde fundamental modes and two overtones, taking into account the angular eigenfunction (see Eq.~7 of Ref.~\cite{Zhu:2023mzv}). 
For m=4, we simultaneously fit $l=4$ and $l=5$ modes, each with the prograde and retrograde fundamental modes and the first overtones. In addition, we add the quadratically driven prograde and retrograde modes to our fit, with frequencies $\omega_{+220\times +220} = 2\omega_{+220}$ and $\omega_{-220\times -220} = 2\omega_{-220}$, respectively. 
We calculate the spherical-spheroidal mixing coefficients for these QQNMs with frequency $2\omega_{+220}$ and $2\omega_{-220}$ using the \texttt{qnm} package~\cite{Stein:2019mop}, solving for the spheroidal harmonics with oblateness parameter $a\omega$. 

We \emph{do not} find that including the quadratically driven mode with frequency $\omega_{+220\times -220} = \omega_{+220}+\omega_{-220}$ improves the fit, and thereby leave this QQNM out of our fitting. 
This mode arises from the coupling between the prograde and retrograde fundamental modes, and its amplitude is likely suppressed due to the small overlap between the radial eigenfunctions.
Further analytical investigation may shed light on the coupling coefficients between these modes~\cite{Ma_in_prep}. 

We perform our fitting procedure with a range of start times, from $10$ M before to $80$ M after the measured peak of $\left|\psi_4\right|$ for the ${l=m=2}$ angular mode, where $M$ is the mass of the remnant.
We extract the complex amplitude of both the driving QNM $A_{+220}$ and driven QQNM $A_{+220\times +220}$ when their amplitudes are most stable with respect to the fitting start time; specifically, we search for a window of $20$~M within the range of the fitting start times where the standard deviation relative to the mean of the fitted amplitude is the smallest~\cite{Baibhav:2023clw,Zhu:2023mzv}. 
We take the mean of the amplitude across this $20$ M window as the fitted amplitude, and the corresponding standard deviation as the uncertainty in the fit. 
We then calculate the quadratic coefficient as $\frac{A_{+220\times+220}}{A_{+220}^2}$, propagating the uncertainties from the fitted amplitudes. The same procedure is used for the retrograde modes, and the above set of steps is carried out independently for each individual simulation.

\section{Results}\label{sec:res}
\subsection{Spin dependence}
We show the measured amplitude and phase of the prograde quadratic coefficient as a function of spin for a wide range of initial conditions in Fig.~\ref{fig:1}, from both the second order code (black) and numerical relativity (colored). 
Both the amplitude and phase of the quadratic coefficients show agreement across all spin to within $\sim 10\%$, with larger deviations occurring at low and high spin. 

At high spin, the resolution from numerical relativity is only marginally sufficient to resolve the nonlinear sourcing of the driven mode.\footnote{We present numerical convergence in the supplemental material. } 
For the outgoing and prograde initial conditions, the retrograde modes are excited with significant amplitudes, and thereby the relative fitting error of the retrograde modes dominates, causing larger uncertainties for the prograde modes. 
This large excitation (as seen by a distant observer) is because the effective radial potential barrier for the counter-rotating perturbation is shallower than that for the co-rotating one~\cite{Chandrasekhar1984}, so the retrograde component of the perturbation near the black hole horizon can escape to distant observers more easily.
For the prograde initial conditions, the retrograde modes are likely introduced from the corrections to the metric arising from solving the constraint equations. 

\begin{figure}[t]%
  \includegraphics[width=\columnwidth]{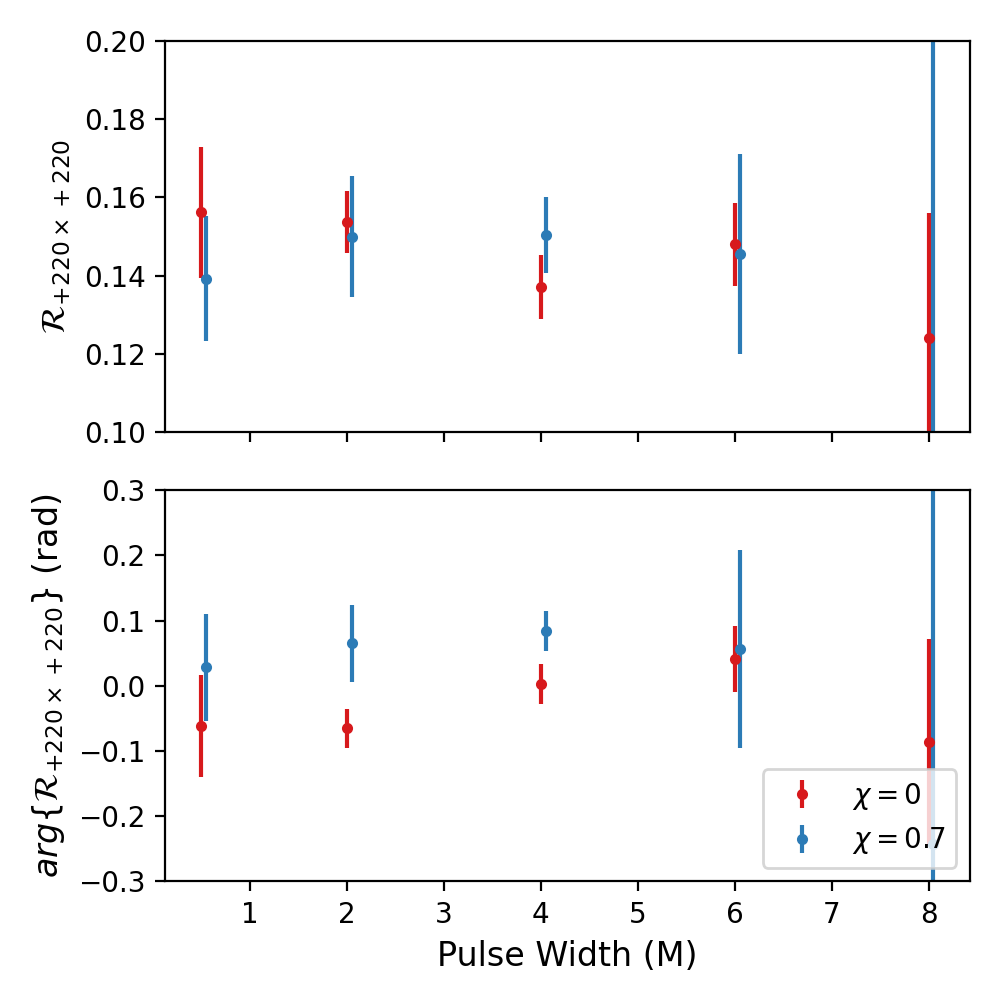}
  \caption{Quadratic coupling coefficient of the prograde ${l=m=2}$ fundamental mode for Schwarzschild and $\chi=0.7$ Kerr black holes as a function of pulse width $w$, calculated from numerical relativity with the ingoing initial condition, with $r_0=30 M$ and the width given by the x axis. The blue points are shifted slightly to the right to better show the error bars.}
  \label{fig:3}
\end{figure}

At low spin, our measurement of the quadratic coupling constant from numerical relativity and the second order code disagree by a margin of $3$ times the fitting errors, while the semi-analytical calculation in Ref.~\cite{Ma_in_prep} seems to be in better agreement with the numerical relativity result, sitting at only $1.5~\sigma$ away.
Here, we give a few possible explanations for the disagreement. 
First, the quadratic coefficient may depend mildly on the initial data, and as explained above we are not at present able to set the same initial data for the second order perturbation and numerical relativity codes.\footnote{In Fig.~\ref{fig:3}, we show results from changing the pulse width in the numerical relativity code. This suggests variations of $\sim10\%$ in the quadratic coefficient with width are possible, however the estimated uncertainties here are of similar magnitude.} 
Second, the stability of our quasinormal mode fits with respect to the fitting start time is likely a necessary but not sufficient criteria for extracting the amplitude of a QNM, so the uncertainties we use could underestimate the true uncertainty of the measured QNM amplitudes~\cite{Berti_personal_communication}.
We cannot rule out the possibility that there is a source of systematic error, from various aspects of theory to numerical evolution codes.
We leave this $10\%$ disagreement between the second order perturbation theory and numerical relativity results as an open question for future investigation. 

Intriguingly, we find that the quadratic coefficient decreases monotonically with black hole spin up to $\chi=0.99$. While the decreasing trend is in general agreement with the findings of Ref.~\cite{Redondo-Yuste:2023seq}, the coupling coefficient from our analysis remains finite at high spin but goes to zero as in Ref.~\cite{Redondo-Yuste:2023seq}. Furthermore, the coupling coefficient in this current work disagree with those reported by Ref.~\cite{Redondo-Yuste:2023seq} by a factor more than two at low spin. 

For near-extremal black holes, one might naively expect that the nonlinear effects are amplified due to the presence of ``zero-damped'' modes, which decay very slowly as the black hole spin approaches extermality~\cite{Aretakis:2012ei,Yang:2014tla,Gralla:2016sxp,Redondo-Yuste:2022czg}. 
This being said, the precise analytical form of the quadratic coupling between QNMs is yet to be understood for Kerr black holes. 
We additionally caution that one should not extrapolate this decrease towards the extremal limit. 
At exact extremality, the frequency of the QQNM coincides with the prograde ${\ell=m=4}$ QNMs, which are now zero-damped with frequencies lying exactly at the onset of superradiance~\cite{Glampedakis:2001js,Yang:2012he,Yang:2013uba}. 
Therefore, one could expect potential nonlinear resonances, and the quadratic coupling coefficient could change drastically towards $\chi=1$. 
Additionally, Ref.~\cite{Yang:2014tla} proposed that a different re-summation is required to study turbulence at near extremal spin, where standard perturbation theory methods may fail. 
We defer a detailed analysis of near-extremal black hole perturbations to future work. 

\begin{figure}[t]%
  \includegraphics[width=\columnwidth]{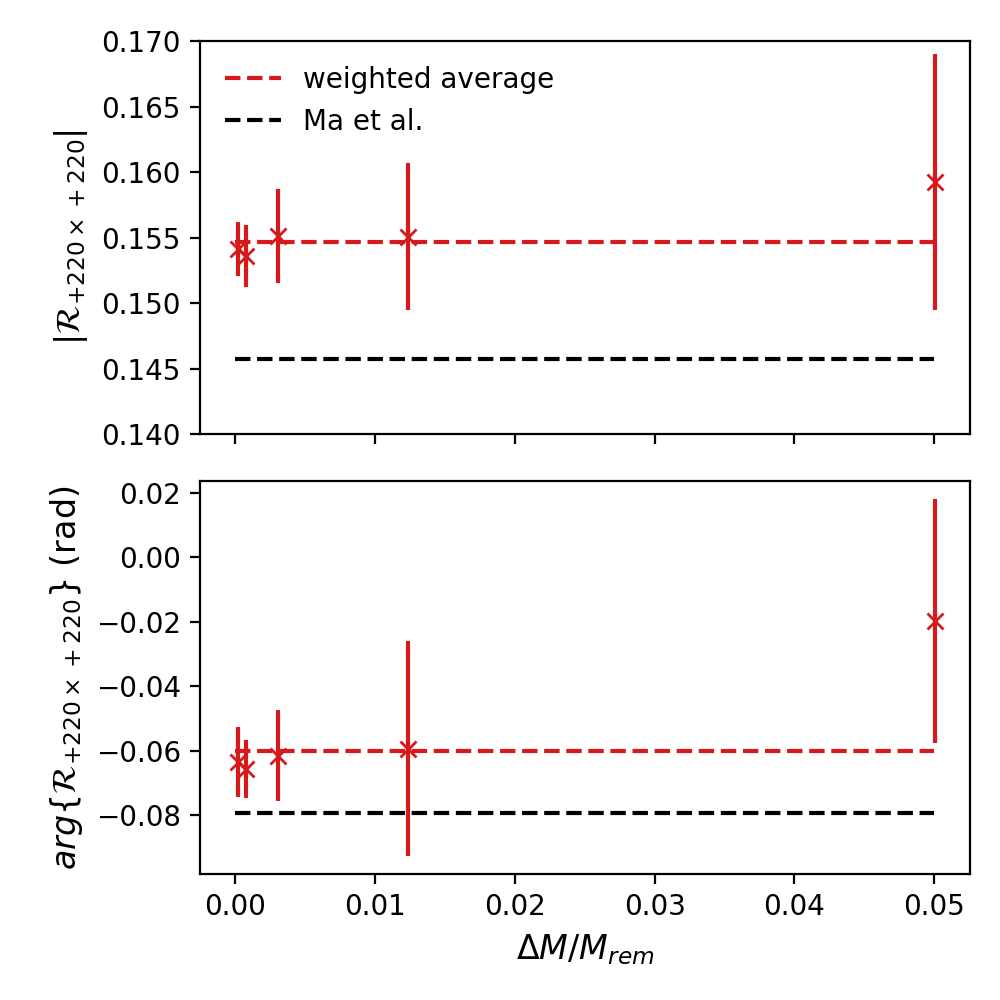}
  \caption{Quadratic coupling coefficient of the prograde ${l=m=2}$ fundamental mode for Schwarzschild vs. the fractional change in black hole Christodoulou mass, calculated from numerical relativity with the ingoing initial condition, with $r_0=30 M$ and $w=2 M$. The red dashed line is the average of the measured values weighted by the uncertainties, and the black dashed line shows the semi-analytical calculation in Ref.~\cite{Ma_in_prep}.}
  \label{fig:2}
\end{figure}

\subsection{Initial data dependence}

In astrophysical mergers we expect the initial perturbation to the remnant black hole to strongly depend on the progenitor binary properties~\cite{Kamaretsos:2012bs,Hamilton:2023znn,Zhu:2023fnf,Bandyopadhyay:2023ohl}. 
Without {\em ab initio} calculations of the spectrum of linear modes excited for a given progenitor, null tests of the second order dynamics of Kerr black holes as predicted by general relativity could still be carried if the quadratic coupling coefficients do not or depend only weakly on initial data.  
In this section, to help assess whether this is the case, we study additional initial data dependence of the quadratic coefficient. 
Beyond the three classes of initial conditions with the range of spin presented in Fig.~\ref{fig:1}, we explore the dependence on the perturbation amplitude (or equivalently higher-order effects) and pulse width for the ingoing pulse initial condition, using the full numerical relativity code.

\emph{Dependence on perturbation amplitude--}
For moderate to equal mass ratio mergers, the remnant black hole mass and spin could change considerably during ringdown due to the back-reaction of the relatively large amount of gravitational energy emitted. 
A pertinent question then is how this may affect the quadratic coupling coefficients.

In Fig.~\ref{fig:2}, we show the quadratic coefficients measured from scattering ingoing pulses of different amplitudes off a Schwarzschild black hole. The largest amplitude results in a change in the black hole mass of roughly 5\%, consistent with what one would expect in an equal mass merger. The measured values all agree within the fitting error, across a large range of perturbation amplitude (by a factor of $\sim 50$ in the energy of the perturbation).
Note that this does not imply that there are no measurable higher order effects present, as we are only fitting for the QQNM, which filters out third and higher order effects that would possess different characteristic frequencies. 
Furthermore, higher order effects would decay faster than the quadratic ones; since our fitting is performed at rather late times (roughly 20 M after the peak) these higher order effects would have already decayed significantly, e.g. absorption-induced mode excitation~\cite{Sberna:2021eui}. 

\emph{Dependence on radial width--} Astrophysically, the characteristic radial scale of the initial perturbation to the remnant black hole is not well-understood and likely also depends significantly on parameters of the progenitor binary, such as the mass ratio and the spin configurations.

In Fig.~\ref{fig:3}, we investigate the dependence of the quadratic coefficient on the width of the ingoing gravitational wave pulse.
We set the width $w$ to range from a quarter to four times the Schwarzschild radius of the background ADM spacetime mass. 
We find that for both Schwarzschild and $\chi=0.7$ Kerr black holes, the quadratic coupling coefficient is rather insensitive to the pulse width, within fitting errors. For the narrower pulse, this error arises from the larger truncation error due to limited resolution. 
For the wider pulses, the error comes from the longer-lasting transient and significant presence of the retrograde modes, affecting the mode extraction.
Nevertheless, the variation is sufficiently small to suggest that the quadratic coupling coefficients could be governed by properties intrinsic to the Kerr background, rather than being functions of the initial data, consistent with arguments from the analytical findings in Ref.~\cite{Perrone:2023jzq,Ma_in_prep}. 
This opens up further possibilities for testing nonlinear dynamics of GR with black hole ringdown spectroscopy.

\begin{figure}[ht!]%
  \includegraphics[width=\columnwidth]{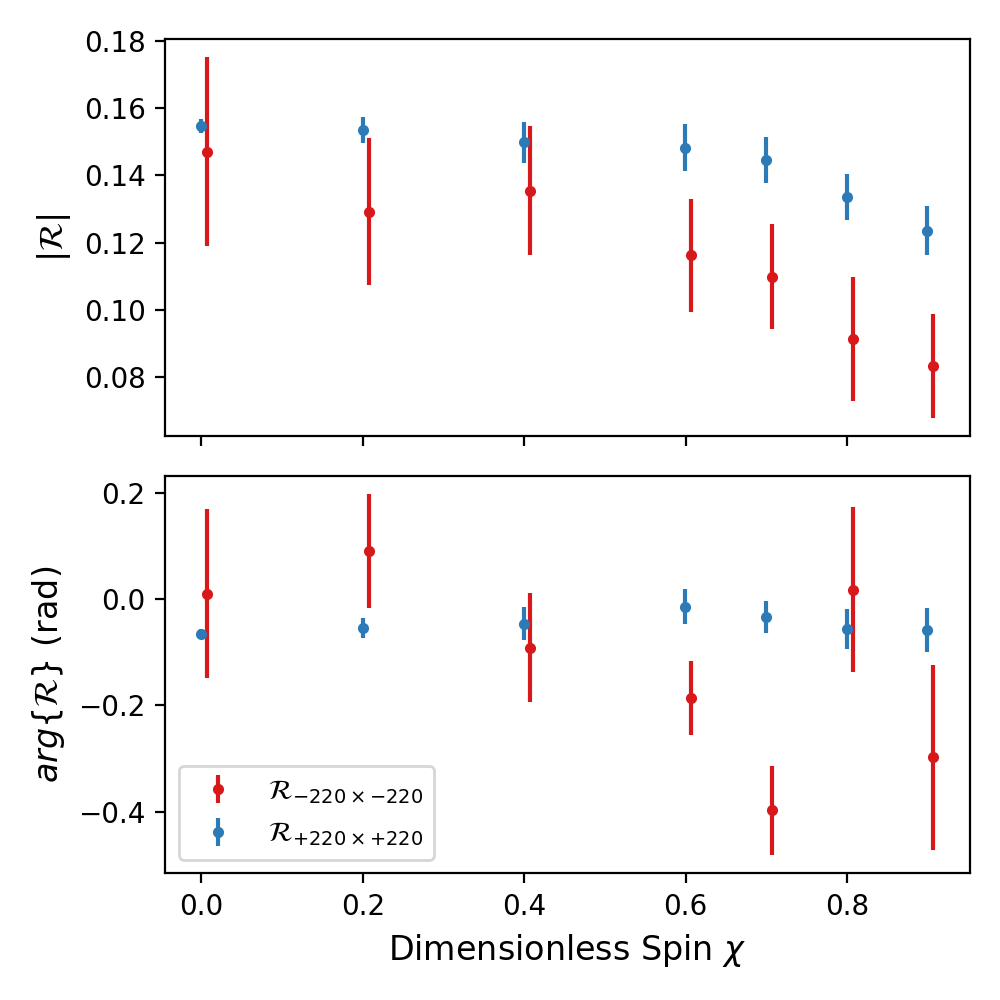}
  \caption{Quadratic coupling coefficient for the prograde and retrograde $l=m=2$ fundamental modes as a function of spin. Both are measured with the ingoing initial condition using numerical relativity. The prograde mode is measured with a pulse width $w=2M$ whereas the retrograde mode with $w=8M$, as wider pulse preferentially excites the retrograde modes (see the supplemental material)~\cite{Andersson:1995zk}. 
  The red dots are shifted slightly to the right to better show their error bars.
  We cannot get a stable estimate of the retrograde coupling coefficient from the second order code because the $1/r$ compactification under-resolves the far-field region, therefore fails to resolve a wide pulse traveling inward away from the black hole horizon. }
  \label{fig:4}
\end{figure}
\subsection{Retrograde mode coupling}
Thus far in the literature, only the coupling between the prograde QNMs has been demonstrated \cite{London:2014cma,Ma:2022wpv,Mitman:2022qdl,Cheung:2022rbm,cheung2023extracting}.
A natural question of theoretical interest is whether the retrograde modes couple with the same strength as the prograde modes. 
In Fig.~\ref{fig:4}, we show the quadratic coupling coefficients among the prograde and retrograde ${\ell=m=2}$ fundamental modes. 
The two coupling coefficients agree for Schwarzschild and bifurcate as the black hole spin increases. 
This is expected as the frequencies and the radial eigenfunctions of the prograde and retrograde modes diverge as the black hole spin increases. 
The prograde modes are more efficiently trapped within the black hole's light ring, leading to lower decay rate towards higher spin.
On the other hand, the retrograde modes are more efficiently absorbed at higher spin, therefore leading to a larger decay rate~\cite{Yang:2012he}.
Further analytical investigations may shed light on the physical origin of this bifurcating behavior~\cite{Ma_in_prep}. 

\section{Conclusion}\label{sec:conclude}
We investigated quadratic coupling of the ${\ell=m=2}$ QNMs during black hole ringdown by simulating a single perturbed black hole. 
Using both numerical relativity and second order perturbation theory, we explored the quadratic coupling coefficient as a function of black hole spin up to $\chi=0.95$ and $0.99$ respectively.
Regardless of the initial condition, we find close alignment between second-order perturbation theory predictions and numerical relativity results, with a discrepancy of less than $10\%$. At moderate to high spins, this difference was within estimated uncertainties, but not so in the limit $\chi=0$. This particular anomaly awaits further exploration in future investigations. 

Furthermore, we found that the quadratic coupling coefficient for the fundamental ${l=m=2}$ modes decreases with increasing spin, by a factor of almost $2$ going from $\chi=0$ to $\chi=0.99$. 
Earlier studies of this coefficient from binary black hole merger simulations that led to remnants with spins $\chi\in0.5-0.7$ are consistent with our results in this range, supporting the conclusion that there is at most mild dependence on initial conditions. 
This implies that quadratic coupling coefficients encode properties more intrinsic to the remnant Kerr black hole than details of the perturbation, and hence could be used to test the non-linear dynamics of Kerr black holes as predicted by general relativity. 

We also performed a preliminary study of the coupling from retrograde QNMs, finding they decrease more rapidly with spin compared to their prograde counterparts. Understanding this in more detail, as well as an investigation of quadratic coupling beyond the dominant quadrupolar modes, is left to future works.

\section{Acknowledgements}\label{sec:acknowledgements}

We thank Emanuele Berti, Alejandro Cárdenas-Avendaño, Mark Cheung, William East, Will Farr, Elena Giorgi, Lam Hui, Maximiliano Isi, Macarena Lagos, Lionel London, Nicholas Loutrel, Harrison Siegel, and Zihan Zhou for helpful discussions regarding various aspects of this project. 
H.Z. especially thank Will Farr, Maximiliano Isi, and Harrison Siegel for hosting stimulating discussions at the Flatiron institute, and Lam Hui and Macarena Lagos at Columbia University, where several of the key ideas presented in this paper were developed. 
The authors are pleased to acknowledge that the work reported on in this paper was substantially performed using the Princeton Research Computing resources at Princeton University which is a consortium of groups led by the Princeton Institute for Computational Science and Engineering (PICSciE) and Office of Information Technology's Research Computing.
Part of the computations for this work were also performed with the Wheeler cluster at Caltech. 
This work was supported in part by the Sherman Fairchild Foundation and NSF Grants, PHY-2011968, PHY-2011961, PHY-2309211,  PHY-2309231, OAC-2209656 at Caltech, and NSF Grants No. PHY-2207342 and OAC-2209655 at Cornell.
J.L.R. acknowledges support from the Simons Foundation through Award number 896696 and the NSF through the award PHY-2207650. F.P. acknowledges support from the NSF through the award PHY-2207286.


\section*{Supplemental Materials}
\appendix
\section{Tetrad Convention}
We review the tetrad normalization conventions and coordinates used in the \texttt{teuk-fortran-2020} second-order code \cite{justin_ripley_2023_8215577} and the \texttt{SpEC} code \cite{Boyle:2019kee}. 
As shown below, to match \texttt{teuk-fortran-2020} with \texttt{SpEC} output at future null infinity, we need to multiply the ratio of the second order curvature perturbation to the square of the first order curvature perturbation--$\Psi_4^{(2)}/\left(\Psi_4^{(1)}\right)^2$--from \texttt{teuk-fortran-2020} by a factor of 2.

\subsection{Classical Kinnersley tetrad}%

In Boyer-Lindquist coordinates, Ref.~\cite{Teukolsky:1973ha} derived the Teukolsky equation using the Kinnersley tetrad \cite{10.1063/1.1664958}, which is given by
\begin{subequations}
\begin{align}
    l_{Kin}^{\mu}\partial_{\mu}
    =&
    \frac{r^2+a^2}{\Delta_{BL}}\partial_t
    +
    \partial_r
    +
    \frac{a}{\Delta_{BL}}\partial_{\phi}
    ,\\
    n_{Kin}^{\mu}\partial_{\mu}
    =&
    \frac{r^2+a^2}{2\Delta_{BL}}\partial_t
    -
    \frac{\Delta_{BL}}{2\Sigma_{BL}}\partial_r
    +
    \frac{a}{2\Sigma_{BL}}\partial_{\phi}
    ,\\
    m_{Kin}^{\mu}
    =&
    \frac{1}{\sqrt{2}\left(r+ia\sin\theta\right)}\left(
        ia\sin\theta\partial_t
        +
        \partial_{\theta}
        +
        \frac{i}{\sin\theta}\partial_{\phi}
    \right)
    .
\end{align}
\end{subequations}
As $r\to\infty$, we have
\begin{subequations}
\begin{align}
    l_{Kin}^{\mu}\partial_{\mu}
    \to&
    \partial_t
    +
    \partial_r
    ,\\
    n_{Kin}^{\mu}\partial_{\mu}
    \to&
    \frac{1}{2}\left(\partial_t - \partial_r\right)
    ,\\
    m_{Kin}^{\mu}\partial_{\mu}
    \to&
    \frac{1}{\sqrt{2}r}\left(
        ia\sin\theta\partial_t
        +
        \partial_{\theta}
        +
        \frac{i}{\sin\theta}\partial_{\phi}
    \right)
    .
\end{align}
\end{subequations}
\subsection{Tetrad use in \texttt{SpEC}}%

From Ref.~\cite{Iozzo:2020jcu}, we see that the tetrad used in \texttt{SpEC} is
\begin{subequations}
\begin{align}
    l_{SpEC}^{\mu}\partial_{\mu}
    =&
    \frac{1}{\sqrt{2}}\left(s^{\mu}\partial_{\mu} + r^{\mu}\partial_{\mu}\right)
    ,\\
    n_{SpEC}^{\mu}\partial_{\mu}
    =&
    \frac{1}{\sqrt{2}}\left(s^{\mu}\partial_{\mu} - r^{\mu}\partial_{\mu}\right)
    ,\\
    m_{SpEC}^{\mu}\partial_{\mu}
    =&
    \frac{1}{\sqrt{2}r}
    \left( 
        \theta^{\mu}\partial_{\mu}
        +
        \frac{i}{\sin\theta}\phi^{\mu}\partial_{\mu}
    \right)
    .
\end{align}
\end{subequations}
where $s^{\mu}$ is a unit timelike vector, $r^{\mu}$ is a unit radial vector, $\theta^{\mu}$ is a unit angular vector, and $\phi^{\mu}$ is a unit angular vector.
As $r\to\infty$, we see that this tetrad reduces to 
\begin{subequations}
\begin{align}
    l_{SpEC}^{\mu}\partial_{\mu}
    \to&
    \frac{1}{\sqrt{2}}\left(\partial_t + \partial_r\right)
    ,\\
    n_{SpEC}^{\mu}\partial_{\mu}
    \to&
    \frac{1}{\sqrt{2}}\left(\partial_t - \partial_r\right)
    ,\\
    m_{SpEC}^{\mu}\partial_{\mu}
    =&
    \frac{1}{\sqrt{2}r}
    \left( 
        \partial_{\theta}
        +
        \frac{i}{\sin\theta}\partial_{\phi}
    \right)
    .
\end{align}
\end{subequations}

\subsection{Tetrad used in \texttt{teuk-fortran-2020}}%
In the \texttt{teuk-fortran-2020} code, we make the following tetrad transformation to the Kinnersley tetrad (for the background) \cite{Ripley:2020xby}
\begin{subequations}
\begin{align}
    l_{TF}^{\mu}
    =&
    \frac{\Delta_{BL}}{2\Sigma_{BL}}l_{Kin}^{\mu}
    ,\\
    n_{TF}^{\mu}
    =&
    \frac{2\Sigma_{BL}}{\Delta_{BL}}n_{Kin}^{\mu}
    ,\\
    m_{TF}^{\mu}
    =&
    \mathrm{exp}\left[
        - 2i \mathrm{arctan}\left(\frac{r}{a\sin\theta}\right)
    \right]
    m_{Kin}^{\mu}
    .
\end{align}
\end{subequations}
As discussed in \cite{Ripley:2020xby}, this tetrad is chosen as it sets the Ricci rotation coefficient $\gamma=0$ on the background, which simplifies the metric reconstruction equations.
Note that for the Kinnersley tetrad, which is the more commonly used tetrad in the GW literature, $\epsilon=0$ instead of $\gamma=0$. 
As $r\to\infty$, we see that
\begin{subequations}
\begin{align}
    l_{TF}^{\mu}
    \to&
    \frac{1}{2}l^{\mu}_{Kin}
    =
    \frac{1}{\sqrt{2}}l^{\mu}_{SpEC}
    ,\\
    n_{TF}^{\mu}
    \to&
    2n^{\mu}_{Kin}
    =
    \sqrt{2} n^{\mu}_{SpEC} 
    ,\\
    m_{TF}^{\mu}
    \to&
    m^{\mu}_{Kin}
    =
    m^{\mu}_{SpEC}
    .
\end{align}
\end{subequations}
The Weyl scalar $\Psi_4$ is
\begin{align}
    \Psi_4
    \equiv
    -
    C_{\mu\nu\alpha\beta}n^{\mu}\bar{m}^{\nu}n^{\alpha}\bar{m}^{\beta}
    .
\end{align}
As $r\to\infty$, we see that
\begin{align}
    \lim_{r\to\infty}\Psi_{4,TF}
    =
    \lim_{r\to\infty}4\Psi_{4,Kin}
    =
    \lim_{r\to\infty}2\Psi_{4,SpEC}
    .
\end{align}
With this, we have
\begin{align}
    \lim_{r\to\infty}\frac{\Psi_4^{(2)}}{\left(\Psi_4^{(1)}\right)^2}
    =
    \frac{1}{4}\lim_{r\to\infty}\frac{\Psi_{4,Kin}^{(2)}}{\left(\Psi_{4,Kin}^{(1)}\right)^2}
    =
    \frac{1}{2}\lim_{r\to\infty}\frac{\Psi_{4,SpEC}^{(2)}}{\left(\Psi_{4,SpEC}^{(1)}\right)^2}
    . 
\end{align}
Therefore, we see that in order to compare at future null infinity (effectively $r\to\infty$ in hyperboloidally compactified coordinates or outgoing null coordinates), we need to multiply the output of the \texttt{teuk-fortran-2020} code by a factor of $2$ comparing to the output of the \texttt{SpEC} code.

\section{Numerical Convergence}

As our setup for the \texttt{teuk-fortran-2020} second-order code is exactly the same as in Ref~\cite{Ripley:2020xby}, we refer interested readers to there for studies of its numerical convergence. Here, we present results from our convergence tests of the full \texttt{SpEC} numerical relativity simulations we ran. 

To simulate a single perturbed black hole in \texttt{SpEC}, we use 48 nested spherical shells extending from inside the outer horizon to 420 M. 
Each domain is integrated on a single MPI rank (cpu thread) using a pseudo-spectral method.

To test the convergence of our simulations, we evolved ingoing initial data for black holes with initial dimensionless spins of $\chi=0$ and $\chi=0.95$, with four different resolutions.
The resolution of a given run (which we label as ``Lev 0-3'') is specified by the number of radial collocation points and maximum L of the angular spherical harmonics expansion for each subdomain.
We list the value for each resolution in Table~\ref{tab:s1}. 
We use the maximum (Lev 3) resolution for the runs presented in the main text.

\begin{center}
\begin{table}[h]
\begin{tabular}{|c || c | c|}
 \hline
 Lev & $N_r$ & $N_L$\\
 \hline\hline
 0 & 10 & 16\\ 
 \hline
 1 & 12 & 18\\
 \hline
 2 & 14 & 20\\
 \hline
 3 & 16 & 22\\
 \hline
\end{tabular}
\caption{Number of collocation points/spectral elements for each resolution. }\label{tab:s1}
\end{table}
\end{center}

\begin{figure*}[htbp]%
    \includegraphics[width=16.cm]{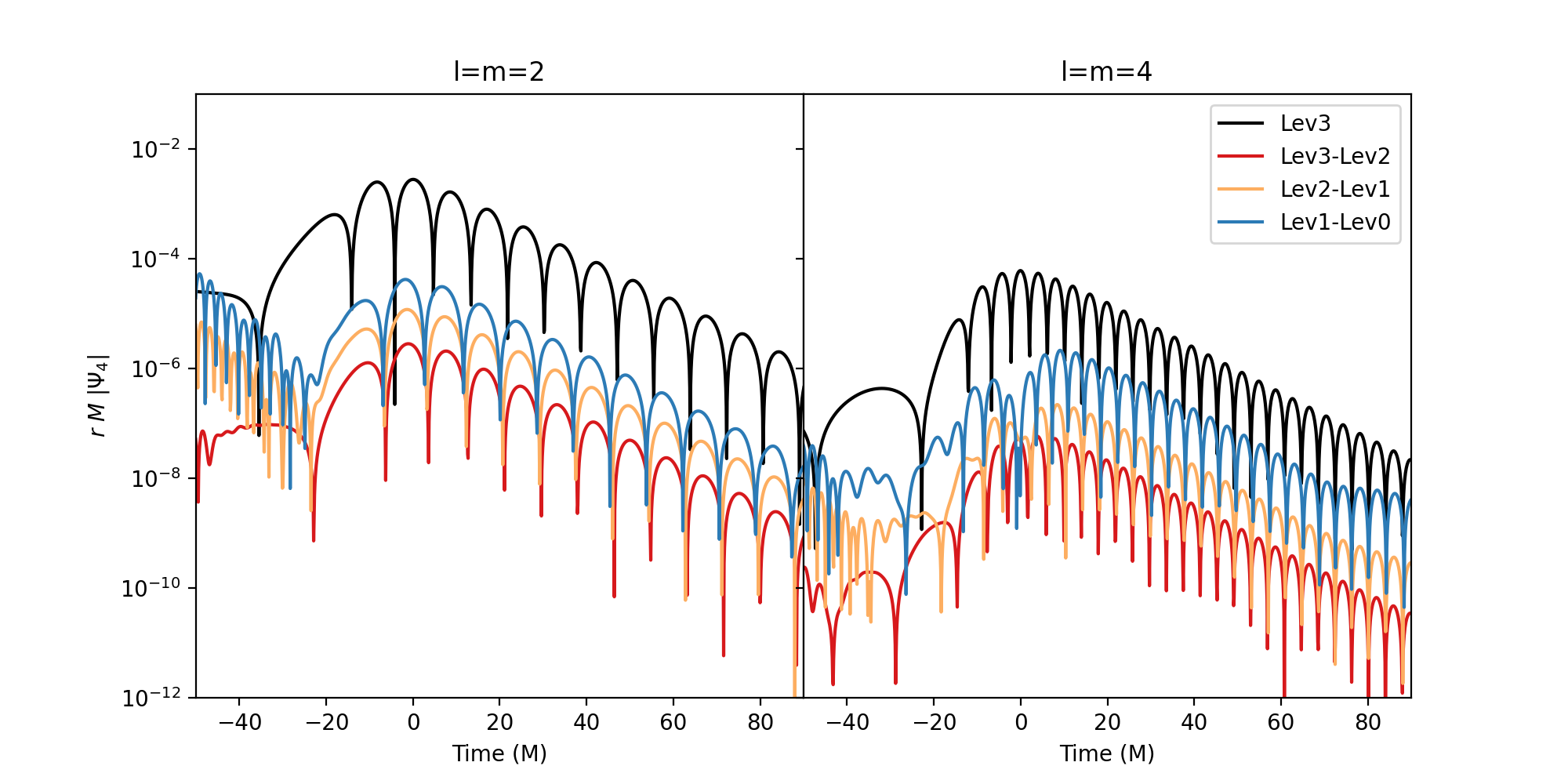}
    \vspace{0.3cm}
    \centering
    \caption{Convergence of the gravitational waveform extracted at a finite radius, for ingoing pulse initial condition with a width of $w=2M$ onto a Schwarzschild black hole. Convergence of the waveform is similar with other classes of initial data. }
    \label{fig:s1}
\end{figure*}

\begin{figure*}[htbp]%
    \includegraphics[width=16.cm]{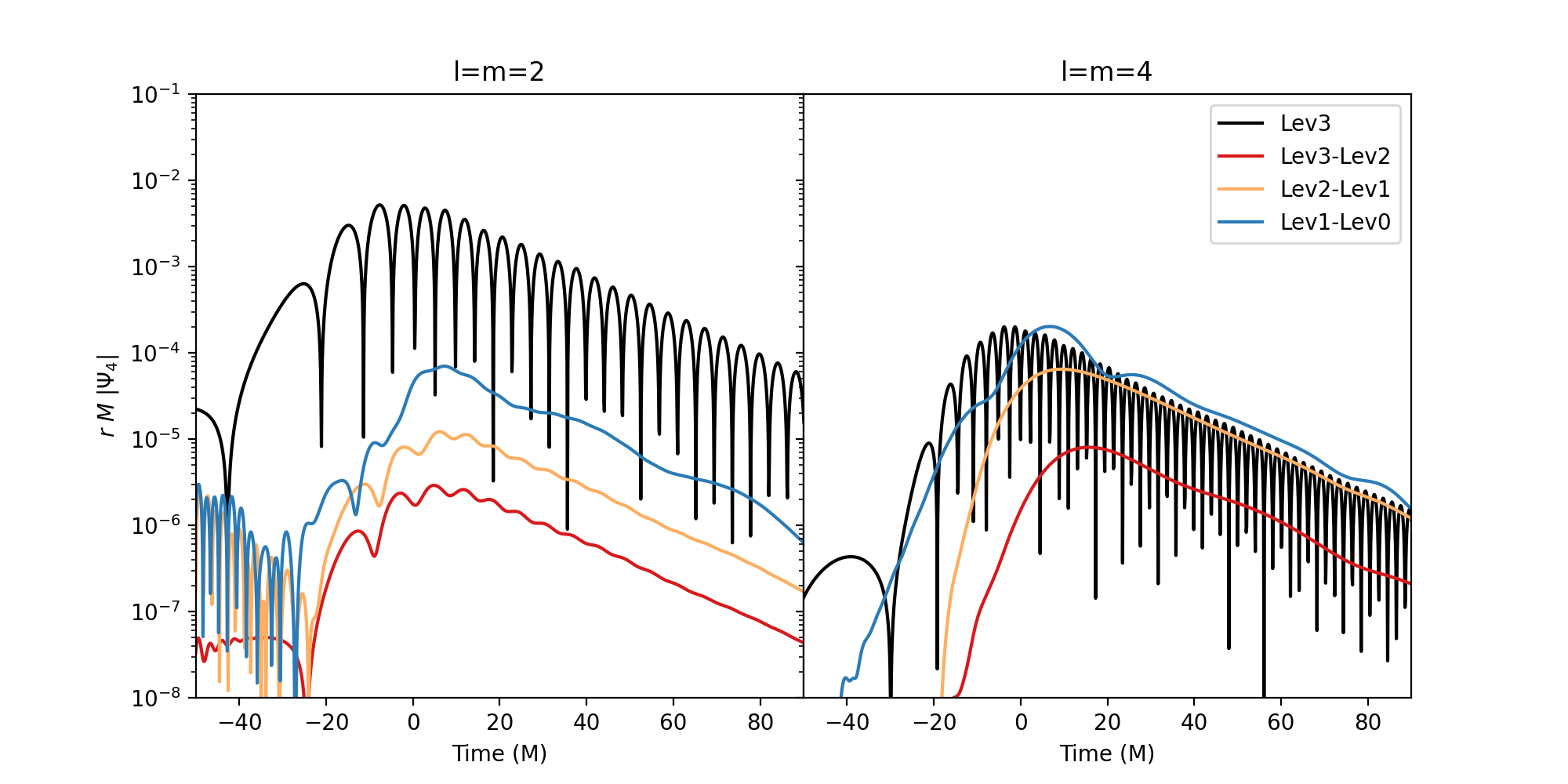}
    \vspace{0.3cm}
    \centering
    \caption{Convergence of the gravitational waveform extracted at a finite radius, for ingoing pulse initial condition with a width of $w=2M$ onto a Kerr black hole with spin $\chi=0.95$.}
    \label{fig:s2}
\end{figure*}

\begin{figure*}[htbp]%
    \includegraphics[width=16.cm]{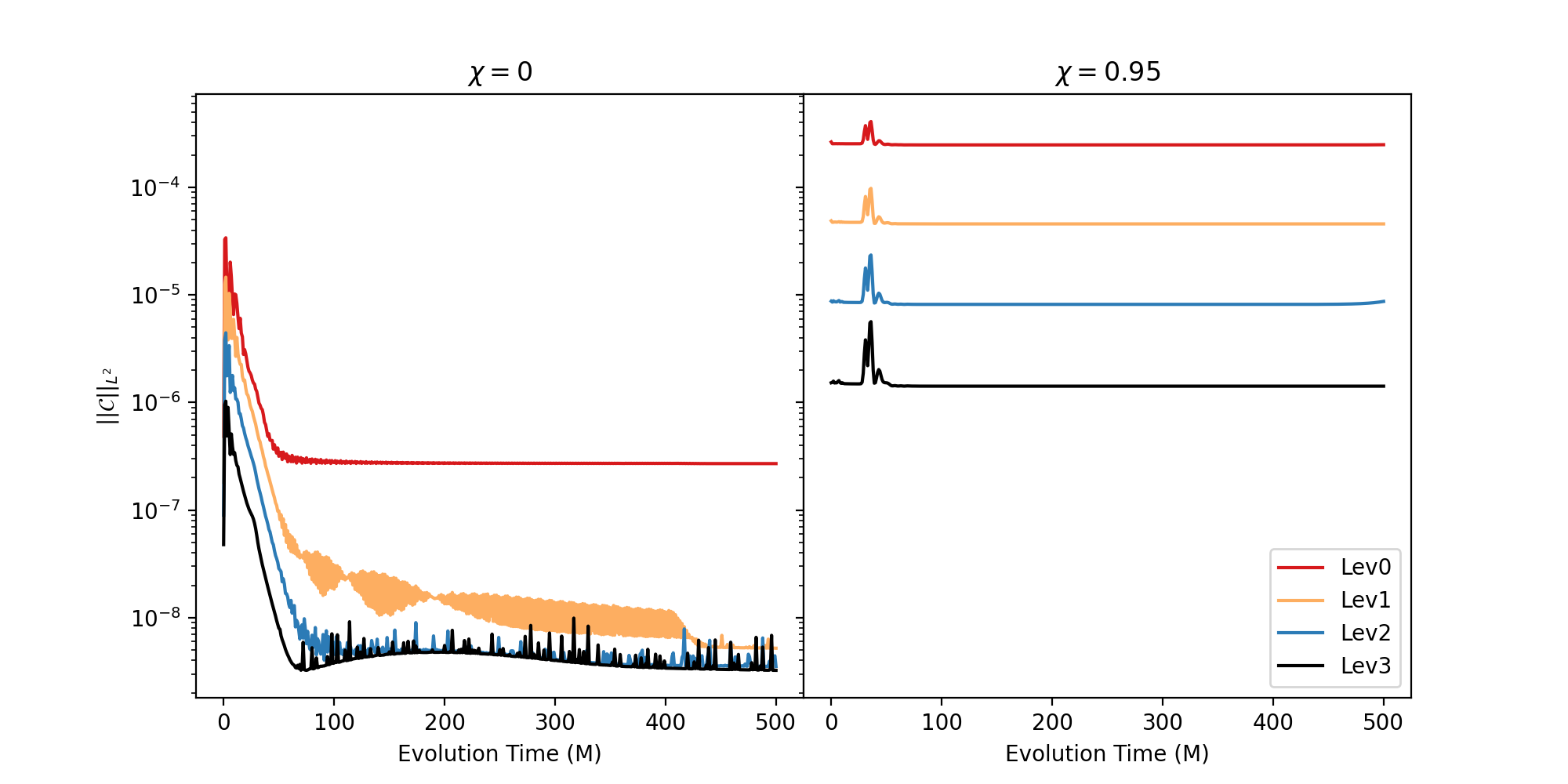}
    \vspace{0.3cm}
    \centering
    \caption{Convergence of constraint violation for $\chi=0$ (left) and $\chi=0.95$ (right). The limiting factor for the $\chi=0$ runs is resolving the ingoing gravitational wave pulse at early time, and round-off error at late time for the higher resolutions, whereas that for the $\chi=0.95$ runs is resolving the background. }
    \label{fig:s3}
\end{figure*}

We show the convergence of a waveform extracted at finite radius for black holes with spins of $\chi=0$ and $\chi=0.95$ respectively in Fig.~\ref{fig:s1}~and~ Fig.~\ref{fig:s2}. 
We show that the quadratic coefficient extracted at finite radius is consistent with that from the CCE waveform in section~\ref{sec:fr_vs_cce}. 
As a final demonstration of the convergence of our numerical relativity simulations, we present the $L^2$ norm of the generalized harmonic constraint (as defined in Eq.~71 in Ref~\cite{Lindblom:2005qh}) for several different resolutions in Fig.~\ref{fig:s3}.

\begin{figure*}[htbp]%
    \includegraphics[width=16.cm]{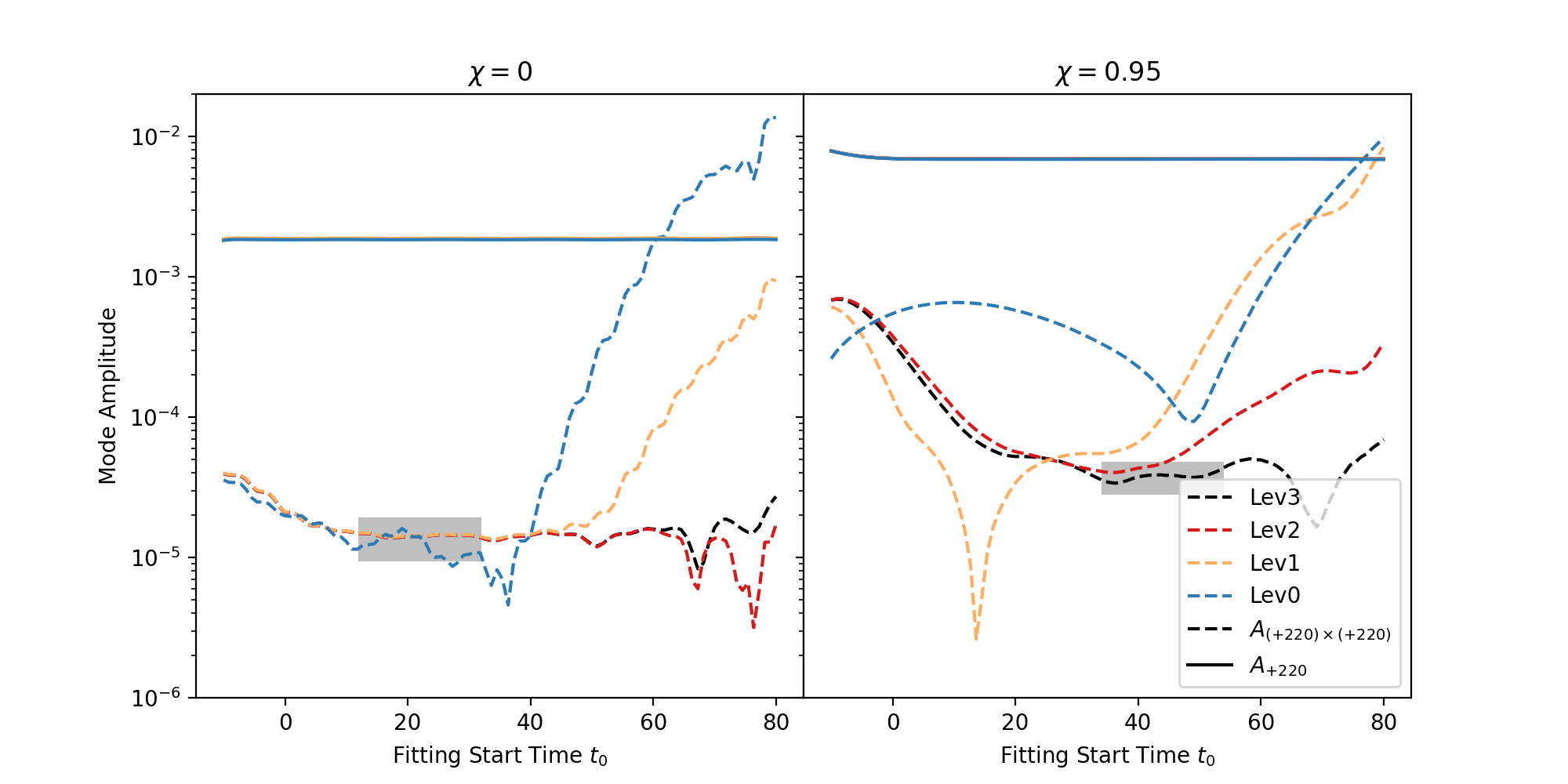}
    \vspace{0.3cm}
    \centering
    \caption{QNM amplitudes from fitting for $\chi=0$ (left) and $\chi=0.95$ (right) with the ingoing initial condition as a function of fitting start time. The solid lines are the prograde $\ell=m=2$ modes, and the dashed lines are the prograde driven QQNM. The resolutions are indicated by color. The grey boxes highlight the most stable intervals in fitting time from which we calculate the mode amplitudes for Lev3 runs.}
    \label{fig:s4}
\end{figure*}

Lastly, we show that our QNM fitting also converges with increasing resolution. In Fig.~\ref{fig:s4}, we present the QNM amplitudes from the fitting as a function of fitting start time, for the prograde $\ell=m=2$ mode and the driven $\ell=m=4$ prograde QQNM. 
We find that the instabilities in the fitting for the QQNM at late time due to numerical error decreases with increasing resolution, and tend to yield a stable amplitude. 
For the Kerr black hole with spin of $\chi=0.95$, the amplitude for the driven QQNM is only marginally stable at the highest resolution. 
Fitting for the retrograde modes \KM{\sout{see} shows} similar convergence.
We note that we \emph{do not} expect the fitting amplitude to converge to a perfect constant line, due to the presence of transients and the tail solution, as discussed in Ref~\cite{Zhu:2023mzv}.

\section{CCE vs. Finite-Radius Waveform}\label{sec:fr_vs_cce}

\begin{figure*}[htbp]%
    \includegraphics[width=16.cm]{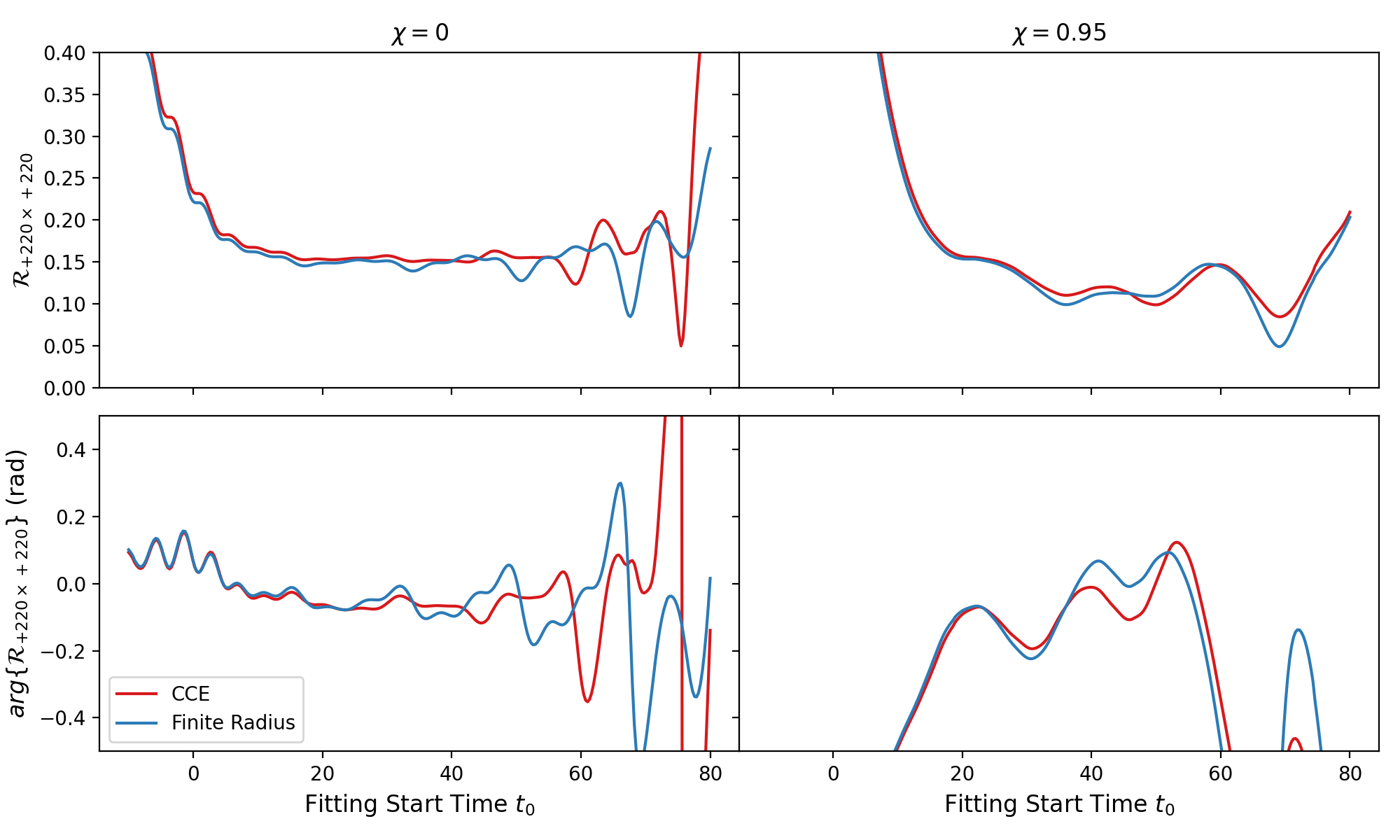}
    \vspace{0.3cm}
    \centering
    \caption{Comparison between the quadratic coefficient extracted from CCE waveform (red) and finite radius waveform (blue) as a function of fitting start time, for $\chi=0$ (left) and $\chi=0.95$ (right). The finite radius waveform is extracted at $R=187M$, and the quadratic coefficients extracted through the two waveforms are consistent within truncation error. $t_0=0$ is the time at peak amplitude of $|\psi_4|$.}
    \label{fig:s5}
\end{figure*}

Finite coordinate/gauge effects may impact one's ability to extract the quadratic coupling coefficient accurately. 
Here, we compare the coefficient inferred from finite-radius waveform extraction and that from Cauchy-characteristic Evolution (CCE). 
In Fig.~\ref{fig:s5}, we plot the coefficient as a function of start fitting time. 
We find that the quadratic coefficients agree within error, yet have slightly different noise characteristics. 
Such good agreement likely stems from our choice of freezing gauge. For binary mergers, where the gauge is more dynamical, the finite radius waveform should be treated with more caution. 

\section{Excitation of Retrograde Modes}
\begin{figure}[ht]%
	\includegraphics[width=\columnwidth]{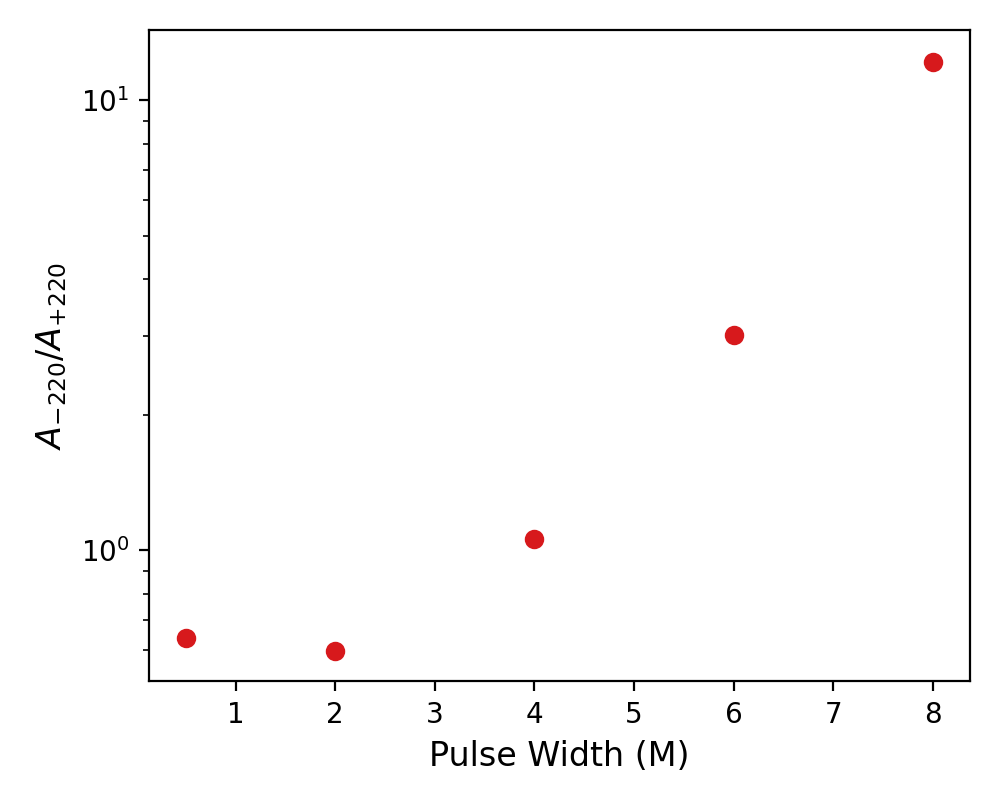}
  \caption{Amplitude ratio between the linearly excited retrograde and prograde fundamental mode as a function of the pulse width for a black hole with $\chi=0.7$. The error bars are too small to be visible.}
  \label{fig:s6}
\end{figure}

To support our finding of the coupling between retrograde modes in the main text, we present additional evidence here.
In Fig.~\ref{fig:s6}, we plot the amplitude ratio between the linearly excited retrograde and prograde mode as a function of the ingoing pulse width on a $\chi=0.7$ Kerr background. 
We find that the relative amplitude of the retrograde mode grows significantly with pulse width, for the range of widths we have tested. At a pulse width of 8 M, the retrograde mode is more than ten times louder than the prograde mode for $\chi=0.7$. Therefore, it becomes the dominant contribution to mode coupling (by a factor of $A^2\sim100$). 

\begin{figure}[ht]%
	\includegraphics[width=\columnwidth]{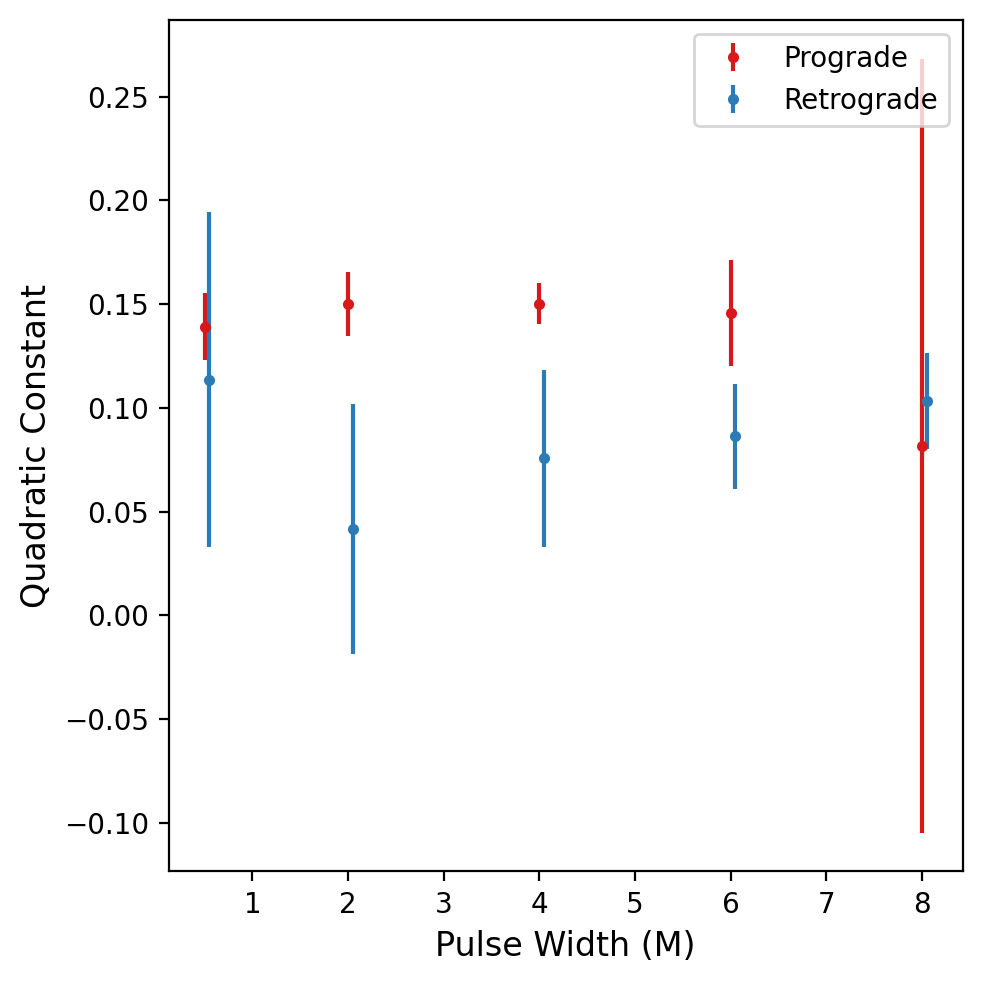}
  \caption{Quadratic coefficient for the prograde (red) and retrograde (blue) mode with different pulse width for a black hole with $\chi=0.7$. We find that the coupling coefficient for the retrograde mode is consistently below that for the prograde mode. }
  \label{fig:s7}
\end{figure}

In Fig.~\ref{fig:s7}, we show that regardless of the width of the ingoing pulse, the quadratic coefficient for the retrograde mode is always below that for the prograde mode for a Kerr black hole with $\chi=0.7$. 
The error on the retrograde coupling coefficient decreases with pulse width, while that for the prograde coupling coefficient increases. 
This is because the fitting algorithm prefers to fit modes with higher amplitude more accurately: the prograde QQNM dominates when the pulse is narrow while the retrograde QQNM dominates when the pulse is wide. QNM filters proposed in Ref~\cite{Ma:2022wpv} could potentially resolve such issue, where one can fit the prograde and retrograde amplitudes separately by filtering out appropriate modes.

\begin{figure}[ht]%
	\includegraphics[width=\columnwidth]{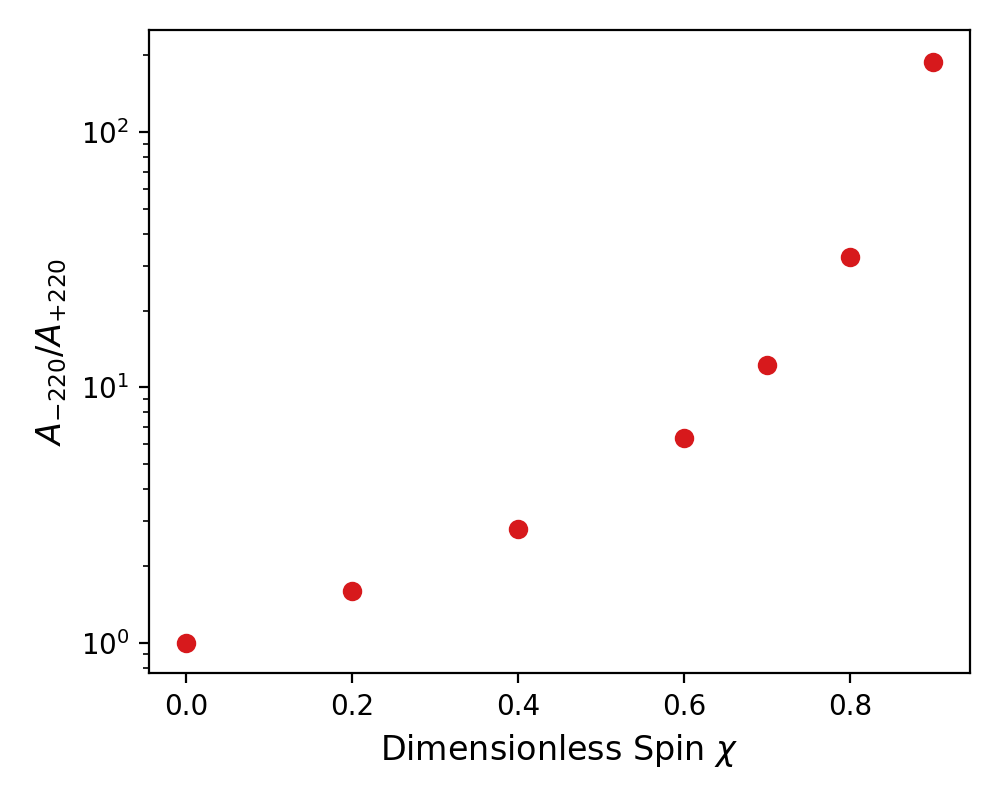}
  \caption{Amplitude ratio between the linearly excited retrograde and prograde fundamental mode as a function of black hole spin, with an ingoing pulse 8 M in width.}
  \label{fig:s8}
\end{figure}

Lastly, in Fig.~\ref{fig:s8}, we further show that with an ingoing pulse width of 8 M, the retrograde amplitude increases further with the black hole spin. 
The physical intuition is that as the spin increases, the length scale of the prograde mode decreases whereas that for the retrograde mode increases. Therefore, a wide pulse would tend to preferentially excite the retrograde content due to the more significant radial overlap.

\clearpage

\bibliography{thebib}
\clearpage
\end{document}